\documentclass[12pt,preprint]{aastex}

\usepackage{amsmath,amssymb,latexsym, graphicx}
\usepackage{epsfig}
\usepackage{booktabs}
\usepackage{afterpage,lscape}
\usepackage[dvips]{color}

\newcommand{\Msun}{~M_\odot}

\newcommand{\msun}{M_\odot}

\newcommand{\gcm}{\rm ~g~cm^{-3}}
\newcommand{\kms}{\rm ~km~s^{-1}}
\newcommand{\ergs}{\rm ~erg~s^{-1}}

\newcommand{\ml}{\Msun ~\rm yr^{-1}}

\slugcomment{Submitted to ApJ}

\shorttitle{Circumstellar interaction in SN 2006jd}
\shortauthors{Chandra et al.}

\begin{document}
\title{RADIO AND X-RAY OBSERVATIONS OF SN 2006jd: ANOTHER STRONGLY INTERACTING TYPE IIn SUPERNOVA}
\author{
Poonam\,Chandra,\altaffilmark{1}
Roger A. Chevalier,\altaffilmark{2}
Nikolai Chugai,\altaffilmark{3}
Claes Fransson,\altaffilmark{4}
Christopher M. Irwin,\altaffilmark{2}
Alicia M. Soderberg,\altaffilmark{5} 
Sayan Chakraborti \altaffilmark{6} \&
Stefan Immler \altaffilmark{7,8,9}
}

\altaffiltext{1}{Department of Physics, Royal Military College of
  Canada, Kingston, ON, K7K 7B4, Canada; Poonam.Chandra@rmc.ca}
\altaffiltext{2}{Department of Astronomy, University of Virginia, P.O. Box 400325, 
Charlottesville, VA 22904-4325.}
\altaffiltext{3}{Institute of Astronomy of Russian Academy of Sciences, Pyatnitskaya St. 48, 
109017 Moscow, Russia.}
\altaffiltext{4}{Department of Astronomy, Stockholm University, AlbaNova, SE-106 91 Stockholm, 
Sweden.}
\altaffiltext{5}{Smithsonian Astrophysical Observatory, 60 Garden St., MS-20, Cambridge, 
MA 02138, USA.}
\altaffiltext{6}{Department of Astronomy and Astrophysics, Tata Institute of Fundamental Research, 
1 Homi Bhabha Road, Colaba, Mumbai 400005, India.}
\altaffiltext{7}{Astrophysics Science Division, NASA Goddard Space Flight Center, Greenbelt, MD 20771}
\altaffiltext{8}{Center for Research and Exploration in Space Science and Technology, NASA Goddard Space Flight Center, Greenbelt, MD 20771}
\altaffiltext{9}{Department of Astronomy, University of Maryland, College Park, MD 20742}

\begin{abstract}
We report four years of radio and X-ray monitoring of  the Type IIn supernova 
SN~2006jd at radio wavelengths with the Very Large Array, Giant Metrewave Radio 
Telescope and Expanded Very Large Array; at X-ray wavelengths 
with {\em Chandra}, {\em XMM-Newton} and {\em Swift}-XRT.
We assume that the radio and X-ray emitting particles are produced by
shock interaction with a dense circumstellar medium.
The radio emission shows an initial rise that can be attributed to
free-free absorption by cool gas mixed into the nonthermal emitting region;
external free-free absorption is disfavored because of the shape of the rising
light curves and the low gas column density inferred along the line of sight
to the emission region.
The X-ray luminosity implies a preshock circumstellar density $\sim 10^6$ cm$^{-3}$ at
a radius $r\sim 2\times 10^{16}$ cm, but the column density inferred from the photoabsorption
of X-rays along the line of sight suggests a significantly lower density.
The implication may be an asymmetry in the interaction.
The X-ray spectrum shows Fe line emission at 6.9 keV that is stronger than is expected
for the conditions in the X-ray emitting gas.
We suggest that cool gas mixed into the hot gas plays a role in the line emission.
Our radio and X-ray data both suggest the 
density profile is flatter than $r^{-2}$ because of the slow evolution
of the unabsorbed emission.
\end{abstract}


\section{INTRODUCTION}
\label{intro}

Type IIn supernovae (SNe IIn) are a  subclass of supernovae (SNe)
that  show characteristic  narrow emission
lines of hydrogen and helium in their spectra  
\citep{sch90,fil97}. These SNe  are thought to
have dense circumstellar media (CSM), and their
high H$\alpha$ and bolometric luminosities
can be explained by the shock
interaction of  SN ejecta with the dense CSM \citep{chu90,cd94}.
Observed late excesses of infrared (IR) radiation are
suggestive of dense circumstellar 
dust \citep{gfn+02,fox11}. 
Generally SNe IIn show significant heterogeneity in terms of their
emission line profiles and luminosities,  and apparent
mass loss history. For example, SN~1998S  demonstrates a connection between
the Type IIL and the Type IIn class: the light curve 
and spectra resemble those of the bright Type IIL SN~1979C 
and yet the H$\alpha$ line shows Type IIn characteristics \citep{fas01}.
SN~1994W faded within 130 days, whereas SN~1988Z remained bright for a decade
after the explosion.  The Type IIn SN~2005gl 
has a possible luminous blue variable (LBV) star 
identified as its progenitor \citep{galyam09}.
SNe IIn are relatively rare,
comprising $\la 10$\% of all core-collapse SNe 
\citep{sma09,li11}.

Dense circumstellar interaction in SNe IIn is expected to 
produce optical, radio and X-ray emission. 
Radio emission in SNe IIn is expected to be  synchrotron emission,
initially absorbed mainly by  free-free absorption, while 
X-ray emission  is likely to have 
a thermal origin \citep{chev82}. Because of the high density of the
CSM, the emission due to the CSM interaction is expected to be high. 
SNe IIn are relatively X-ray luminous, exceeding $10^{40}$ erg s$^{-1}$
in some cases, e.g.,
 SN 1988Z \citep{ft96,sp06} and SN 1986J \citep{bp92,bp94,trs05}.
However, SNe IIn have been elusive at  radio wavelengths. 
So far only 10
SNe IIn have been detected in the radio. 
\citet{vws+96} carried out a 
search for radio emission from 10 SNe IIn and did not detect any.

SN 2006jd was discovered on  2006 Oct 12, with an apparent magnitude of
17.2,  in the galaxy UGC 4179
at a redshift of $z=   0.0186$ \citep{pl06}. The initial spectrum of
SN 2006jd showed features of a Type IIb SN, similar to SN 1993J \citep{bmk+06}. 
However, based on  Keck spectra at late epochs, it was
reclassified as a Type IIn SN
(Alex Filippenko, priv. comm.). 
On the basis of a spectrum on 2006 Oct 17.51 UT, we
assume its explosion date to be 2006 Oct 06.5 UT \citep{bmk+06}.
It is located at R.A. $=08h02m07.43s$, Decl. $=+00^o 48\arcmin 31\arcsec.5$ 
(equinox 2000.0),
which is $22\arcsec$ east and $1\arcsec.3$ south of the nucleus of UGC 4179.
\citet{ifp07} detected  X-ray emission from SN 2006jd
on 2007 Nov 16  with the \emph{Swift} X-ray Telescope (XRT) in a 2.3 ks exposure. The 
net count rate was $(6.3\pm2.0)\times10^{-3}$ cts s$^{-1}$. 
For an adopted plasma model with
a temperature of 10 keV and 
a Galactic column density of $4.5\times10^{20}$ cm$^{-2}$, they
estimated an unabsorbed luminosity of 
$(2.5\pm0.8)\times10^{41}\ergs$, placing it in the category of 
highly X-ray luminous SNe IIn.
Radio emission was detected from the SN on 2007 Nov 21.28 UT in the 5 GHz
band, on 2007 Nov 26.36 UT in the 8.5 GHz band, and on 2007 Nov 26.38 UT
in the 22.5 GHz band \citep{cs07}.

In this paper, we report 4 years of radio and X-ray monitoring of SN 2006jd and its
implications for the CSM density and the mass loss rate of the
progenitor star. In Section \ref{obs}, we describe the radio and X-ray observations,
and their analysis in Section \ref{datanalysis}.
We discuss the main results and present
 conclusions in Section \ref{discussion}.

\section{OBSERVATIONS}
\label{obs}

\subsection{Radio Observations}
\label{Radio}

Radio observations of SN 2006jd started on 2007 Nov 21.28 UT with the
Very Large Array \footnote{The
Very Large Array is operated by the National Radio Astronomy
Observatory, a facility of the National Science Foundation operated
under cooperative agreement by Associated Universities, Inc.} (VLA).
We observed the SN in the 22.5 GHz (K band), 8.5 GHz (X), 5 GHz (C), 
and 1.4 GHz (L) bands at various epochs,
along with a 44 GHz band observation at one epoch.
For C and X bands, the data were taken in the interferometric mode
for an average of 30 minutes (including calibrator time), whereas for L and K bands 
the data were collected for 
1 hr.  The total bandwidth used was 50 MHz. 3C48 was used 
as the flux calibrator and J0739+016 was used for the phase calibration. The VLA
observed SN 2006jd over the period 2007 Nov 21.28 to  2009 Aug 7.75. From  2010 Jun onwards, the 
observations were made using the Expanded VLA (EVLA). For EVLA observations,
3C48 and J0739+016 were also used as  bandpass calibrators. The data were
analyzed using standard AIPS routines.

We also observed SN 2006jd with the Giant Meterwave Radio Telescope (GMRT)
on three occasions between 
2009 Oct  to 2010 Apr  in the 1.3 GHz and 0.61 GHz bands. 
We detected the
SN in the 1.3 GHz band at both occasions, whereas 
observations at 0.61 GHz resulted in upper limits.
For GMRT observations, 3C48 and 3C147 were used as flux calibrators and 
J0739+016 as a phase calibrator. All these calibrators 
were also used for bandpass
calibration. In the GMRT observations, 
the observing time on SN 2006jd was approximately
3 hours in each observation and a bandwidth of 32 MHz was used. 
AIPS was used to analyze the
GMRT data as well.

Details of all the radio observations are listed in Table~\ref{tab:radio}. 
We show the radio light curves of SN 2006jd at various frequencies in 
Figure~\ref{radio-lc}
and radio spectra at various epochs in Figure~\ref{radio-spectra}.

\subsection{X-ray observations}
\label{xray}

SN 2006jd was observed with the \emph{Swift} onboard X-ray Telescope (XRT) 
at various epochs between
2007 Nov  and 2011 Mar. In addition, 
we also observed it once with   the \emph{Chandra} and
once with the  \emph{XMM-Newton} X-ray observatories. 

We carried out the \emph{XMM-Newton} observations 
under our proposal \# 55085 starting
2009 Apr 07 at 4:36:46 UT and  continuing until 16:50:06 UT on
2009 Apr 07. The observations were carried out with the  
EPIC-PN and EPIC-MOS cameras in full frame 
with
thin filter mode. The exposures for the EPIC-MOS1 and
EPIC-MOS2  were 42.667 ks and 42.672 ks, respectively, and
for the EPIC-PN, the exposure time was 41.032 ks. 
The \emph{XMM-Newton} on-board
Optical Monitor (OM) also observed the SN in UVW1, UVM2 and UVW2 filters for
5, 10 and 14.3 ks, respectively. 

We used the EPIC-PN observation to carry out
a detailed spectroscopic analysis. The {\em XMM-Newton}
 software SAS was used to extract the spectrum. 
A total of 1963 counts were obtained in the $0.2-10$~keV range, 
 resulting in a net count rate of 
$(6.19\pm0.17)\times10^{-2}$ cts s$^{-1}$.
The NASA HEASARC software 
HEAsoft\footnote{http://heasarc.gsfc.nasa.gov/docs/software/lheasoft/} 
was used to do the spectral analysis.

We also observed SN 2006jd with  \emph{Chandra} under our proposal
\# 10500688. The observations were 
carried out using ACIS-S with grating NONE in the
VFAINT mode. The observations started on 2009 Sept 14 at 00:03:21 UT for a total exposure
of 37.24 ks. A total of 888 counts were obtained in the $0.2-10$~keV
range with a count rate of $(2.38\pm0.08)\times10^{-2}$ cts s$^{-1}$.
We extracted the spectrum using CIAO software and
used  HEAsoft to analyse the spectrum.

{\em Swift}-XRT observed SN 2006jd on 18 occasions between 2007 Nov  and
2011 Mar. 
All the \emph{Swift} observations were for exposures less than 10 ks.
The observations were carried out in the photon-counting (PC) mode.
To extract the counts from the \emph{Swift} observations, we combined the 
near simultaneous
observations and estimated the count rates using XIMAGE and SOSTA 
of the HEAsoft package.

Table~\ref{tab:xray} gives all the X-ray observations used in this
paper. 
Table \ref{tab:swift} lists the {\em Swift} observations 
which were combined to obtain the 
count rates. The table gives the best estimates of count rates
extracted within the $0.2-10$ keV range.  We also obtained the
best fit positions of SN 2006jd in various {\em Swift} observations and estimated the 
position error in arcseconds from that of the actual SN position.

\section{DATA ANALYSIS AND INTERPRETATION}
\label{datanalysis}
\subsection{Radio data analysis and interpretation}
 
\begin{figure*}
\centering
\includegraphics[width=0.98\textwidth]{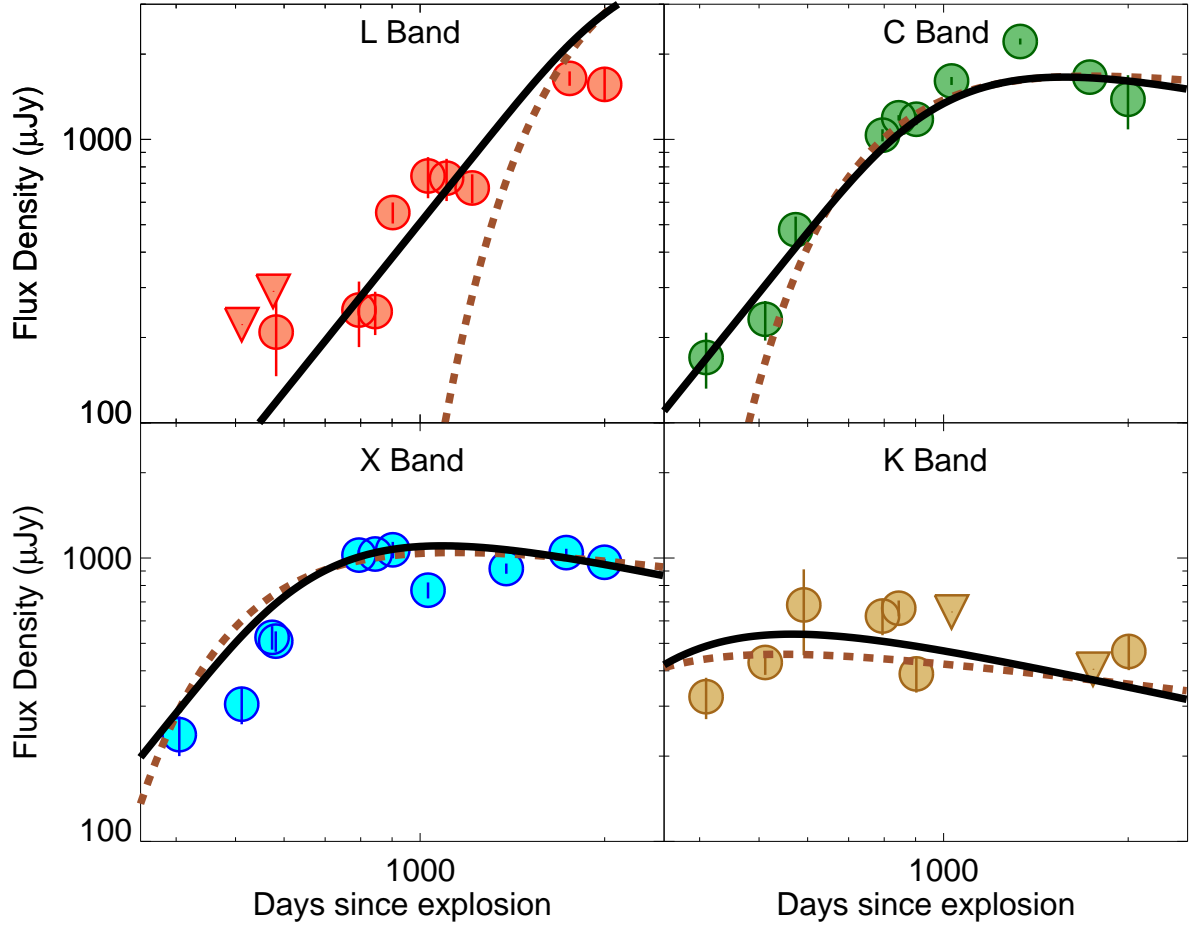}
\caption{Radio light curves of SN 2006 in the 1.4 GHz (L),
5 GHz (C), 8.5 GHz (X) and 22.5 GHz (K) bands.
The radio emission in the L band is still rising and the K band emission
appears to be in the 
optically thin phase at late times. The radio emission is making a 
transition from optically thick to thin  in the X and C bands.
 Here dashed lines show the best fit external absorption model for $s=1.77$.
 Black solid lines are the best fit internal absorption model.}
\label{radio-lc}
\end{figure*}

\begin{figure*}
\centering
\includegraphics[width=0.94\textwidth]{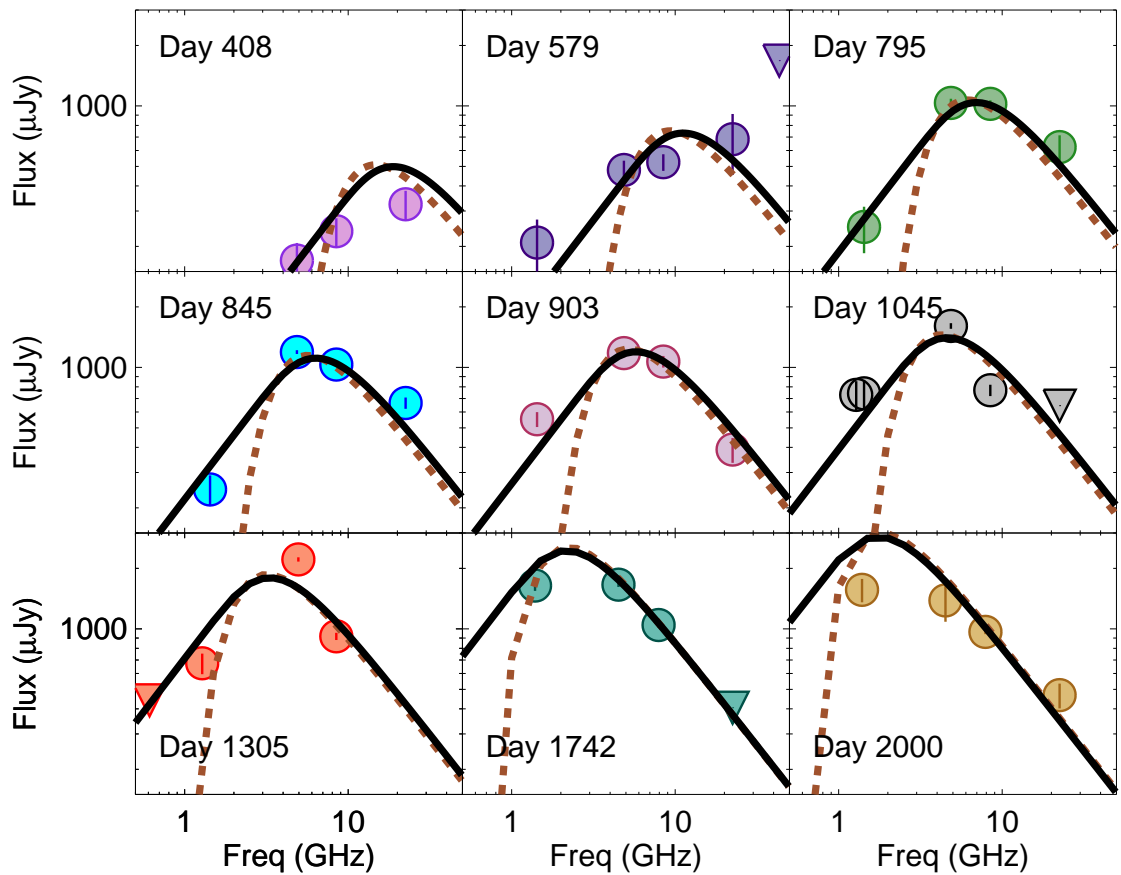}
\caption{Radio spectra of SN 2006jd at 9 epochs. Even though the
spectral evolution is slow, a transition from
optically thick  to an optically thin  is
clearly seen in the progression  of the spectra from
 early  to  late epochs. Dashed lines show the best fit external absorption model for $s=1.77$.
 Black solid lines are the best fit internal absorption model.
}
\label{radio-spectra}
\end{figure*}

\begin{figure*}
\centering
\includegraphics[width=0.98\textwidth]{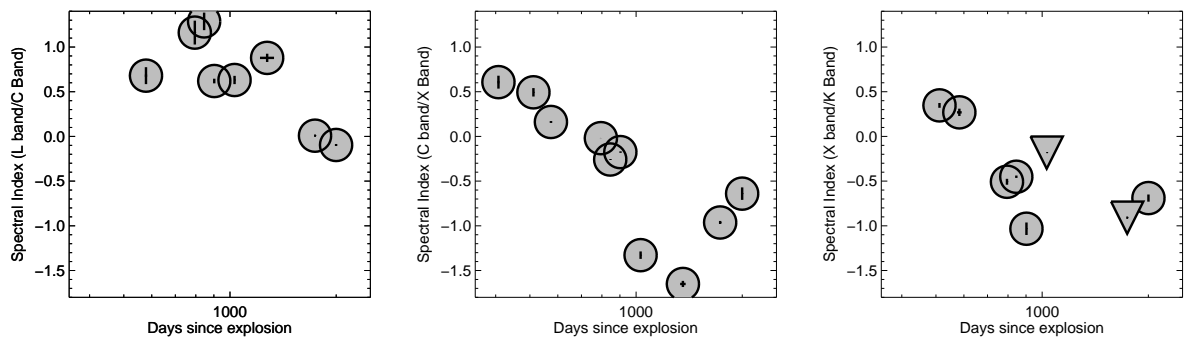}
\caption{Spectral index ($\alpha$) evolution for SN 2006jd between
various radio frequencies. The evolution is from positive index  towards
negative, typical in core-collapse SNe.}
\label{alpha}
\end{figure*}

SN 2006jd  remained bright in all the radio bands over the 4 year span
of observations, except for two early upper limits in L band and occasional upper limits in the
 K band at later epochs. Figure~\ref{radio-lc} shows the light curve of the SN in L, C, X and K bands.
The SN was in the optically thick regime in the L band, whereas
it  reached the optically thin regime in the K band.
In the X and the C bands, the SN was close to the peak of the light curve, 
indicating a transition from the optically thick to the optically thin phase.
The near-simultaneous spectra are plotted from day 408 to day 2000 (see Figure~\ref{radio-spectra}).
The radio observations of SN 2006jd show an evolution 
from a somewhat positive spectral index
to a  negative index (see Figure~\ref{alpha}).
This type of evolution is commonly observed  
in radio SNe \citep{weiler02} and is attributed
to the transition from optically thick to optically thin radiation.
The  transition can be clearly seen by
comparing progressive spectra from the earlier to the late epochs, where the peak is
 shifting from the higher frequencies to lower frequencies with
time.

The primary mechanisms for absorption in the radio 
emission are free-free absorption by 
the circumstellar material (FFA) and synchrotron
self-absorption \citep[SSA,][]{chev82,cf03}.
The relevance of SSA can be determined from the peak radio 
luminosity and the time
of peak, which can be related to the velocity 
of the radio emitting region if SSA
dominates \citep{chev98,cf03}.
For SN 2006jd, we find a velocity of $2000-3000\kms$ assuming the SSA absorption
mechanism, 
 which is smaller than  
expected in a typical SN and also than the velocities deduced from the X-ray emission
for SN 2006jd (Section \ref{xrayanalysis}).
We thus conclude that SSA cannot account for sufficient
absorption and FFA is  likely to dominate, although
more complicated models (e.g., CSM clumpiness or global asymmetry)
might relax constraints on the SSA mechanism imposed by 
the expansion velocity. 

We  concentrate on FFA models,
allowing for the possibility of
a relatively flat density profile in SN 2006jd.
At the high mass loss rates characteristic of SNe IIn, it is plausible
that the density profile is not that of a steady wind.
We fit the external FFA model \citep{chev82,weiler02} to
the SN 2006jd data with a generalized CSM  power law density 
dependence  $\rho\propto r^{-s}$ \citep{fransson96}:
\begin{eqnarray}
F(\nu,t)=K_1 
\left(\frac{\nu}{5 \,{\rm GHz}}\right)^{\alpha}
\left(\frac{t}{1000\, {\rm day}}
\right)^{\beta} 
\exp(-\tau_{\rm FFA}), \nonumber \\ 
\tau_{\rm FFA}=K_2 \left(\frac{\nu}{5 \,{\rm GHz}}\right)^{-2.1}
\left(\frac{t}{1000\, {\rm day}} \right)^{\delta}, 
\label{eq:ffa}
\end{eqnarray}
where $\alpha$ is the optically thin frequency spectral index, which relates to the electron
energy index $\gamma$ ($N(E) \propto E^{-\gamma}$) as
$\gamma=1-2 \alpha$.  To determine the value of $\alpha$,
we choose the average of spectral indices between
K and X band  and X and C band in Figure~\ref{alpha} after it has become optically thin.  
This value is $\alpha=-1.04\pm0.05$. Thus we fix $\alpha=-1.04$  
($\gamma=3.08$) in the radio absorption models. The parameter $\delta$ is related to the
expansion parameter $m$ ($R \propto t^{m}$, where $R$ is the shock radius) and the
CSM density power law index $s$ (in $\rho\propto r^{-s}$) as $\delta=m (1-2s)$.
The assumption that the energy density in the
particles and the fields is proportional to the postshock energy density
leads to $\beta=3m-(3-\alpha)(ms+2-2m)/2$.
Here $K_1$ is the radio flux density normalization
parameter and $K_2$ is the external FFA optical depth normalization parameter.
Using these parameters, we fit the SN 2006jd radio data to Eq. \ref{eq:ffa}.
We first fit the data for standard $s=2$ model in Eq. \ref{eq:ffa}, which gives a
very poor fit (Table \ref{tab:radiofit}). The best fit $\delta$ here is $\delta=3.46\pm0.02$
implying $m=1.15$. Allowing for a flatter density profile, 
our data is best fit with a value $m=1.09$ and $s=1.77$  (Table  \ref{tab:radiofit}).
Dashed lines in Figures~\ref{radio-lc} and \ref{radio-spectra}
show the external FFA fit to the radio data for SN 2006jd with $s=1.77$.  
The fit to the data is not perfect and the value $m>1$ is a problem because
the model for the emission requires that $m\le 1$ \citep{chev82}.
We explored models with $m<1$, but found that they did not give an acceptable
fit to the data.

A property of the external absorption model is that there is an initial
exponential ($e^{-\tau}$) rise in the flux.
The data appear to give more of a power law rise, which is also observed in other
Type IIn supernova where the rise of the radio flux was followed:
SN 1986J \citep{weiler90} and
SN 1988Z \citep{vd93,williams02}.
\cite{weiler90} proposed a model in which thermal absorbing gas is mixed into
the synchrotron emitting gas, so that the flux takes the form
\begin{eqnarray}
F(\nu,t)=K_1 
\left(\frac{\nu}{5 \,{\rm GHz}}\right)^{\alpha}
\left(\frac{t}{1000\, {\rm day}}
\right)^{\beta} 
\left(  \frac{1-\exp(-\tau_{\rm intFFA})}{\tau_{\rm intFFA}}\right), \nonumber \\ 
\tau_{\rm intFFA}=K_3 \left(\frac{\nu}{5 \,{\rm GHz}}\right)^{-2.1}
\left(\frac{t}{1000\, {\rm day}} \right)^{\delta^{\prime}}, 
\label{eq:int}
\end{eqnarray}
where again we allow $s\ne 2$, so that $\beta$ can be
expressed as before.
\cite{weiler90} assume that the internal absorbing gas is homologously expanding
with density $n\propto R^{-3}\propto t^{-3m}$.
Then $\tau_{\rm intFFA}\propto n^2R\propto t^{-5m}$, or $\delta^{\prime}=-5m$.
We did not make that assumption in our fits because there are other possibilities for
the evolution of the density.
The internal absorbing gas is likely to be in pressure equilibrium with the surrounding
hot gas and photoionized by the energetic radiation field.
If the gas temperature evolves slowly, we have $n\propto p \propto R^{2-s}t^{-2}$ where
$p$ is the pressure, so
that $\delta^{\prime}=-4+m(5-2s)$.
In our fits, we allow $\delta^{\prime}$ to be a free parameter.

We considered models that allowed for external as well as internal FFA.
However, we found that external absorption played a negligible role in the
best fit case, while the internal absorption by itself provided an excellent fit.
The fits are shown in Figures~\ref{radio-lc} and \ref{radio-spectra} and the model
 parameters are shown in Table \ref{tab:radiofit}.
A notable feature of the fits is the small magnitude of $\beta$, which is 
associated with the slow evolution in the optically thin regime that can be
seen in the late light curve evolution (Figure  \ref{radio-lc}).
Taking $s=2$ for the given $\beta$ implies $m=1.19$, which is not consistent
with the hydrodynamic model.
However, a density profile with $s=1.5$ gives $m=0.9$, which is a
plausible value.
In this model, the value of $\delta^{\prime}$ is between the values found in the
two density evolution models described above.

It can be seen from equation (\ref{eq:int}) that in the optically thick limit
$F(\nu)\propto \nu^{2.1+\alpha}$.  
The value of $\alpha$ is determined by the late optically thin evolution.
The fact that the optically thick spectrum is approximately reproduced (Figure  \ref{radio-spectra})
provides support for the internal FAA model.

The preferred model has an optical depth of unity at 5 GHz at an age of 1000 days.
Assuming that the absorbing gas is in pressure equilibrium with the X-ray
emitting gas and that it has a relatively low temperature ($10^4-10^5$ K),
it is possible to estimate the mass of the gas.
We determined the X-ray emitting gas has a temperature $T>20$ keV.
If we take $T\approx 60$ keV, corresponding to a shock velocity $v_{sh}=6700\kms$,
the density of the X-ray emitting gas is $6\times 10^6$ cm$^{-3}$ at 1000 days (Section 3.3).
In pressure equilibrium, the density of cool gas is then $n_c\approx 3.6\times 10^{11}T_4^{-1}$
cm$^{-3}$, where $T_4$ is the temperature in units of $10^4$ K.
The optical depth is $\tau = \kappa_{\nu}\ell$, where $\ell$ is the path length through the
cool gas and $\kappa_{\nu}=3.6\times 10^{-27}n^2 T_4^{-3/2}$ cm$^{-1}$ is the free-free
absorption coefficient.
The absorbing mass is $M_a\approx 4\pi R_c^2\bar mn_c\ell$, where $R_c$ is the typical
radius of the absorbing gas, $\bar m$ is the mean particle weight, and the optical depth
constraint can be used to determine $n_c\ell$.
The result is $M_a\approx 2\times 10^{-8}T_4^{5/2}\Msun$, showing that a modest amount of cool
gas mixed into the emitting region can give rise to the needed absorption.
The source of the cool gas is likely to radiative cooling of dense gas in the shocked
region.

The models presented here assume a constant spectral index for
the radiating electrons.
Physical effects, such as Coulomb losses at low energies and synchrotron
losses at high energies, can affect the particle spectrum \citep{fransson98}.
Estimates of these effects indicate that they might be significant, but a
complete discussion is beyond the aims of this paper.

\subsection{X-ray data analysis}

\subsubsection{XMM-Newton}

\begin{figure}
\centering
\includegraphics[angle=-90,width=0.48\textwidth]{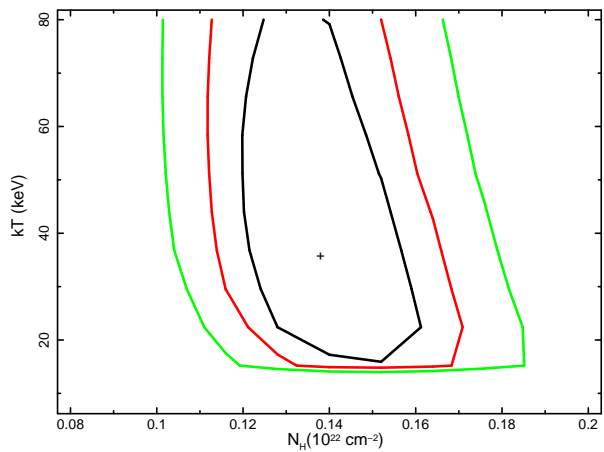}
\caption{Confidence contours of column density versus the 
X-ray emitting plasma temperature. 
The column density is well constrained whereas the temperature is not 
constrained at the higher end.}
\label{cc-nh-kt}
\end{figure}

The X-ray data were best fit with a thermal plasma model at a temperature above 10 keV,
i.e. in a range where the $0.2-10$ keV spectral shape is not sensitive to temperature.
The spectrum also indicates a  
column density of $1.3 \times10^{21}$ cm$^{-2}$, 
which is significantly larger than
the Galactic absorption column density 
of $4.5\times 10^{20}$ cm$^{-2}$ in that direction.
To determine the significance of the best fit temperature and the
column density, we plot 68\%, 90\% and 99\% confidence contours of
column density $N_H$ versus the plasma temperature $kT$. While the
column density is a well constrained parameter, the
temperature is unconstrained at the upper end (see Figure~\ref{cc-nh-kt}). 
To better constrain the temperature, 
we lowered the temperature until the $\chi^2$ became significantly
worse and the fit
was visibly bad. This temperature is 20 keV. We thus assign 20 keV as the
lower limit to the plasma temperature. 

\begin{figure}
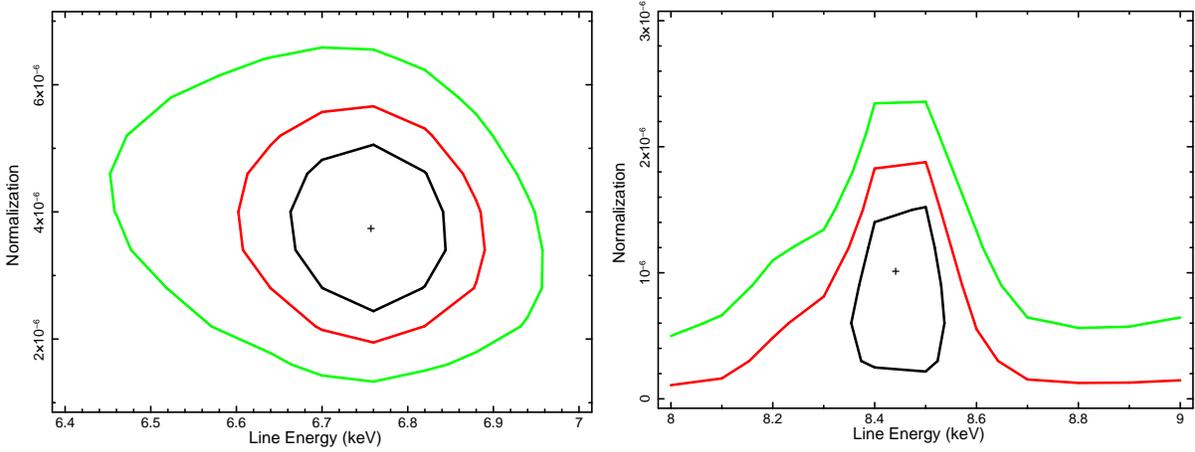

\centering
\includegraphics[angle=-90,width=0.48\textwidth]{f5a.eps}
\includegraphics[angle=-90,width=0.48\textwidth]{f5b.eps}
\caption{Confidence contours of the Fe 6.9 keV line (left panel)
and the 8.1 keV line (right panel).  The iron line is detected at a 
high significance level whereas detection of the 8.1 keV line is not significant above
1-sigma level.}
\label{cc-Fe}
\end{figure}

We have a  clear detection of a line between $6.5-7.0$ keV.
The line is best fit at energy 6.76~keV, which, after redshift
correction, corresponds to an energy of 6.89~keV. We associate this
with the 6.9~keV Fe line, which is due to Fe XXVI.
There is also an evidence of a possible
8.1 keV line, which can be attributed to Ni XXVIII.
To determine the significance of these lines, we plot confidence
contours of their energies versus the respective strengths (Fig~\ref{cc-Fe}). 
The Fe 6.9 keV line is detected at a high significance level, whereas
the 8.1 keV line is significant only at the
68\% level.

\begin{figure}
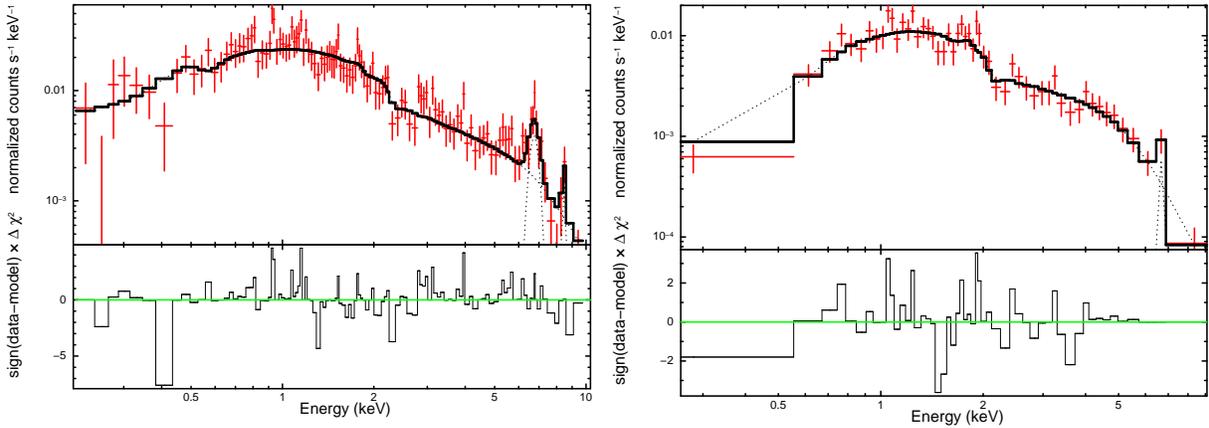

\centering
\includegraphics[angle=270,width=0.48\textwidth]{f6a.eps}
\includegraphics[angle=270,width=0.48\textwidth]{f6b.eps}
\caption{The left panel shows the \emph{XMM-Newton} spectrum of SN 2006jd and
the right panel shows the \emph{Chandra} spectrum of SN 2006jd. Note that the
6.9 keV Fe line is detected in both spectra and has 10\% of the total 
continuum flux (estimated from the {\em XMM-Newton} spectra).
}
\label{xray-spectra}
\end{figure}

Now we attempt to fit the whole spectrum.
Since the temperature is very high, we explore the possiblity of
non-thermal emission. The best fit power law photon index of 
$\Gamma=1.22\pm0.09$ is so flat that we consider this model unlikely. 
If we fit the Mekal model with a solar abundance, it does not reproduce 
the Fe 6.9 keV line.   
However, a Mekal model with 5 times the solar abundance 
does fit the spectrum well and reproduces the iron line at the
correct energy but with a narrower width. 
To look for the possibility of non-equilibrium ionization (NEI),
we also fit the XSPEC NEI model to our data. The fit is quite good and
it reproduces the iron line at the correct energy. 
The best fit ionization time scale in the NEI model
is $\tau=6.28\times 10^{11}$ s cm$^{-3}$. Since $\tau=n t$ ($n$ is the
number density and $t$ is the age) and the XMM spectrum
was taken when the SN age was 908 days (or $7.84\times10^7$ s), this 
implies a number density of $n=7.7\times 10^3$ cm$^{-3}$.
This seems  low  and  we will
discuss the physical viability of the NEI model in  Section \ref{discussion}. 
Finally, we  fit the thermal bremsstrahlung model with one Gaussian
{\rm line fit}
as well as two Gaussians. The fit is not improved by adding the Gaussian at
8.1 keV. Best fits are given in Table \ref{tab:xspec-xmm} and our best fit
spectra are in Figure~\ref{xray-spectra}.  We chose the bremsstrahlung model along with
the Gaussian to be the best fit representation to our data.
The equivalent width of the Fe line is
1.4 keV and the flux in the iron line is about 10\% of the total
continuum flux in the $0.2-10$ keV spectrum.

\subsubsection{Chandra-ACIS}

The {\em Chandra} spectrum is best fit with a high temperature thermal
plasma and a column density of $1.6\times10^{21}$ cm$^{-2}$,
which agree with our fits to the \emph{XMM-Newton} spectrum  very well.
Even though the \emph{Chandra} spectrum is not as 
detailed as that of the \emph{XMM-Newton},
we clearly detect the 6.9 keV Fe line, though the 8.1 keV  line possibly
seen in the
{\em XMM-Newton} spectrum is not present. As in the {\em XMM-Newton} spectrum, the temperature
is not well constrained on the upper end, whereas the column density is
well constrained. The iron 6.9 keV line is also well constrained at the
90\% confidence level.

Because of the high temperature, we again look for the possibility of 
non-thermal emission; a fit yields a power law photon index of 
$\Gamma=1.31\pm0.14$, which again is implausibly flat. The non-equilibrium ionization
model correctly reproduced the Fe 6.9 keV line but again indicates a low density
($n=4\times10^3$ cm$^{-3}$). 
The Mekal model with 5 times solar abundance does 
reproduce
the Fe 6.9 keV line. 
We fit a bremsstrahlung model with a gaussian at 6.9 keV. 
The details of the \emph{Chandra} best fit spectrum are in 
Table \ref{tab:xspec-chandra}.
Fig. \ref{xray-spectra} shows the \emph{XMM-Newton} and \emph{Chandra} spectra
of SN 2006jd.

\subsubsection{Swift-XRT}

The exposures in the \emph{Swift} observations were not large and carrying out an
independent spectroscopic analysis was not possible with the
small number of  counts obtained. Thus
we converted the count rates to $0.2-10$ keV fluxes using a thermal plasma model
with a temperature of 60 keV and column density of
$1.3\times10^{21}$ cm$^{-2}$.

\begin{figure}
\centering
\includegraphics[width=0.48\textwidth]{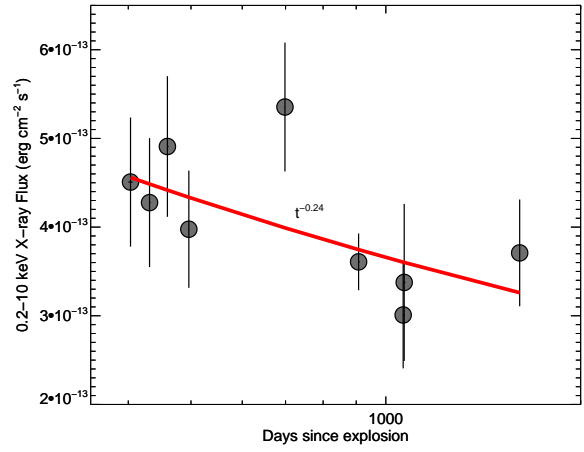}
\caption{X-ray light curve of SN 2006jd. The light curve is
best fit with a power law of index $-0.24\pm0.12$.
}
\label{xray-lc}
\end{figure}

Table \ref{tab:flux} gives the unabsorbed fluxes at the various epochs of
observation
used in this paper. 
We also convert the X-ray fluxes to luminosities using a UGC 4179 
distance of 79 Mpc.
Figure~\ref{xray-lc} shows the $0.2-10$ keV X-ray light curve for SN 2006jd.
The flux evolution is  slow and is best fit with a power law
index of $t^{-0.24\pm0.12}$.

\subsection{X-ray Interpretation}
\label{xrayanalysis}

The {\em XMM-Newton} and
{\em Chandra} X-ray spectra can be fit by either an optically 
thin thermal spectrum, which leads
to an electron temperature $T_e\ga 20$ keV, or a nonthermal spectrum with photon
index $\Gamma\approx 1.2$.
In the nonthermal interpretation, flux $\propto\nu ^{-0.2}$ (as 
$\alpha=1-\Gamma$) would be a
surprisingly hard spectrum for either synchrotron or inverse Compton emission.
In the case of synchrotron radiation, one can show that for a plausible emitting
volume (determined by the supernova shock velocity), the radiating particle would
rapidly lose energy to synchrotron radiation, tending to steepen the spectrum and 
making the flat spectrum more unlikely. Thus we 
discard the nonthermal origin of X-rays and proceed with the
interpretation for the thermal origin of the X-ray spectrum.

In the thermal interpretation, 
the temperature $T$ is assumed to be produced by shock 
heating, with the 
shock velocity $v_{sh}=[16 k T/(3\mu m_p)]^{1/2}=6700(kT/60{\rm~keV})^{1/2}$
km s$^{-1}$, where $k$ is Boltzmann's constant,
 $\mu$ is the mean molecular weight, and $m_p$ is the proton mass.
The shock velocity is closer to expectations for the 
forward shock than the reverse shock wave
at this age,
although X-ray emission from young supernovae is typically attributed to the
reverse shock wave \citep{cf03}.
A high velocity reverse shock wave is possible if 
the SN runs into a dense medium
which acts as a wall for the expanding supernova gas; such a situation is observed
in SN 1987A.
However, this situation is expected to produce rising fluxes 
at radio and X-ray wavelengths
followed by a turnover.
The fluxes in SN 2006jd are observed to decline slowly 
from days 400 through 1700, so we
consider a forward shock origin to be more likely.

We found that the temperature of the emitting region is higher than
can be measured in the $0.2-10$ keV bandpass of the X-ray detectors,
so that a spectral luminosity $L_{\nu}$ is measured rather than a total X-ray luminosity.
For bremsstrahlung emission we have $L_{\nu}\propto n^2V/T^{1/2}$, where $V$ is the
emitting volume, so that $L_{\nu}\propto  t^{2m(1-s)+1}$.
If the observed $L_{\nu}\propto t^{-x}$, then $x=0.24$ for the present case (Figure~\ref{xray-lc}).
We have $x=2m(s-1)-1$ or $s=1+(1+x)/2m=1+0.62/m$.  Taking $m=0.9$ gives $s=1.7$.
The radio and X-ray observations thus give $s\approx 1.6$, which is similar to
the value deduced for SN 2010jl from observations at a earlier epoch \citep{chandra12}.
Using results from \cite{fransson96} for $L_{\nu}$ as in \cite{chandra12}
the mass loss rate $\dot M$, normalized to $R=10^{15}$ cm,
is $\dot M_{-3}/v_{w2}\approx 5v_4^{0.6}$, where $\dot M_{-3}=\dot M/(10^{-3}\ml)$,
$v_{w2}$ is the preshock wind velocity in units of $100\kms$,
and $v_4$ is the average velocity in units of $10^4\kms$ at $10^3$ days.
At $10^3$ days, the preshock density is $2\times 10^{-18} v_4^{-1}\gcm$.
The corresponding density in the shocked gas is $n_H\sim 3\times 10^6$ cm$^{-3}$.

The observed X-ray spectrum is well fit by a single 
temperature model, and there is
no sign of emission from a reverse shock region.
One possible reason for the lack of such emission 
is that the supernova interaction has evolved to the 
point where the reverse shock wave has moved back to the center and has weakened.
In the model described above, the mass swept up by the forward shock is
$M_{sw}=\dot MR_s/v_w\simeq3\msun$.
If the ejecta mass is significantly 
smaller than this, the reverse shock region is expected to
decline in luminosity.
Another possibility is that the reverse shock is a 
cooling shock and the cool shell absorbs the radiation
from the reverse shock region.
In addition, when the reverse shock is radiative, 
the luminosity from the reverse shock region
rises as $\rho_w$, while that from the forward shock region rises as $\rho_w^2$.
In SN 1993J, a hot (80 keV) thermal X-ray 
spectrum was observed in the first few weeks and
interpreted as emission from the forward shock region, with emission from the reverse
shock absorbed by a cooling shell \citep{leising94,fransson96}.
The conditions for radiative cooling at the reverse shock 
depend on the density profile of
the ejecta, and there is a transition to nonradiative evolution at late times.
At the high densities inferred for SN 2006jd, that transition can be delayed.

Here we have assumed that the hot gas is in thermal equilibrium.
The time for the electrons and protons to come into equilibrium is
 $t_{ep}\sim 3$ yr for $kT=50$ keV and $n_e=10^6$ cm$^{-3}$ 
\citep{spitzer62}.
At an age of 1000 days, the electrons and protons
 are thus close to equilibrium by Coulomb
interactions and may be brought even closer to equilibrium by 
plasma instabilities.
If equilibrium has not been obtained, the shock velocity is somewhat 
higher than what we have deduced here.

A robust result of fitting the X-ray spectrum is the hydrogen
 column density $N_H$ to the
source.
Of the $N_H=1.3\times 10^{21}$ cm$^{-2}$ that is observed, $4.5\times 10^{20}$ cm$^{-2}$ can
be attributed to the Galaxy, so there is $8.5\times 10^{20}$ cm$^{-2}$ 
left for the supernova
host galaxy interstellar medium and the supernova circumstellar medium.
This column density estimate assumes 
that the absorbing medium has solar abundances, 
and is cool and not highly ionized.
In the model for the X-ray emission discussed above, 
the expected column density in the
unshocked circumstellar medium is $4\times 10^{22}$ cm$^{-2}$, 
a factor of fifty larger than that inferred
from the observations.
A possible reason is that the circumstellar medium 
is fully ionized by the X-ray radiation.
However, the ionization parameter for this case 
is $\xi=L/nr^2=50$, which indicates that there
may be some ionization of CNO elements, but not full ionization 
of the gas \citep{vf96}.
To check on this result, we ran the CLOUDY code \citep{fer98}
for the parameters discussed above.
As expected, we found some ionization of CNO elements, 
but little change in the absorption
of the X-ray spectrum.

The temperature of the circumstellar medium is 
relevant to the absorption of the radio
emission from the shocked region.  
In the CLOUDY simulation we found an electron 
temperature somewhat less than $10^5$ K,
which is consistent with the ionization parameter.
Using equation (20) of \cite{cf03} and the 
circumstellar parameters derived from the
X-ray luminosity, the wind optical depth at 5 GHz on day 1000 is
$\tau_{ff}=2.4T_{e5}^{-3/2}$, where $T_{e5}$ is 
the electron temperature in units of $10^5$ K.
The optical depth is thus consistent with 
that indicated by the observations
 (Figure \ref{radio-spectra}).
The radio emission is expected to be from 
the high temperature X-ray emitting
region and we have seen that the absorbing column 
to the X-ray emitting region is relatively
small.
Assuming that the inner radius is like that 
found for the X-rays and scaling to the lower column
density (upper limit) leads to $\tau_{ff}=8\times 10^{-4}T_{e5}^{-3/2}$ at 5 GHz on day 1000.
The high ionization parameter in this case 
gives $T_{e5}>1$, so free-free absorption
along the line of sight to the X-ray emission fails to 
produce the observed radio absorption.
This is consistent with our analysis of the radio emission, which showed that external
free-free absorption does not lead to good model fits to the data.

Another robust result from the X-ray spectrum is the finding of an Fe emission line
feature centered at 6.9 keV rest energy (using $z=0.0186$),
 with an equivalent width of 1.4 keV.
The collisional equilibrium model that fits the continuum emission 
has a weak Fe line (0.1 keV equivalent width) at
an energy of 6.9 keV.
The model line is much weaker than that observed.
We considered the possibility of ionization 
non-equilibrium by using the nonequilibrium
model in XSPEC.
The model depends on the postshock temperature, 
which was  fit with 60 keV, and the
ionization time $\tau_i=nt$, which we varied for various fits.
We found that for $\tau_i=6\times 10^{11}$ cm$^{-3}$ s the model provided a good
fit to the data, with $\chi^2/dof=1.03$; 
both the equivalent width and the line energy were well
fit.
However, considering the age of 1000 days, 
this implies the density to be $\sim 7\times 10^3$ cm$^{-3}$ (Section \ref{xray}),
much less than the density in the hot gas, $n_H \sim 3\times 10^6v_4^{-1}$ cm$^{-3}$, needed
to produce the X-ray luminosity.
Raising $\tau_i$ in the nonequilibrium model, we found that by $\tau_i=5\times 10^{12}$
cm$^{-3}$ s, the line was essentially the same as in equilibrium, showing that any density
close to that needed for the X-ray luminosity fails to produce the Fe line.
The nonequilibrium model in XSPEC assumes a Sedov blast wave model for the hydrodynamics,
which is different from the situation studied here, but we do not expect a significant
change in the results.

The equivalent width of K$\alpha$ line
thus exceeds values predicted by ionization equilibrium models. 
A scenario  that might resolve the issue involves the presence of the 
cool dense shell that is formed from cooled gas at the reverse shock wave.
The shell is subject to the Rayleigh-Taylor (RT) instability 
so that the forward postshock zone turns into  a
two-phase mixture composed of rarefied hot and dense cool 
gas \citep{cb95,be01}. The observed 
X-ray continuum with a weak K$\alpha$ line presumably arises 
from the hot component. The two-phase mixture, however, could enhance 
the production rate of K$\alpha$ photons because of the efficient 
recombination of hot Fe$^{+26}$ ions penetrating the cool phase with  
low temperature electrons. This can be illustrated as follows.
The total recombination rate in the $k$-th phase ($k$ is `cool' or `hot') 
is $R_k\propto p_k\alpha_kn_{e,k}$, where $p_k$ is the probability for 
the fast Fe$^{+26}$ ion to get into $k$-th phase, 
$\alpha_k\propto T_{e,k}^{-\eta}$ is the recombination coefficient, and
$n_{e,k}$ is the electron number density. 
Assuming $p_k$ to be proportional to the volume filling factor of the 
corresponding phase (which requires strong mixing),
we find that the ratio of recombination rates of fast Fe$^{+26}$ ions
in the cool and hot phases is 
$R_{c,h}=(M_c/M_h)(T_h/T_{c,eff})^{\eta}$, where $M_k$ is the mass of the corresponding
phase and $T_{c,eff}=\mbox{max}[T_c, T_h(m_e/56m_p)]$ is the effective 
temperature that determines the relative velocity of cool electrons and hot 
Fe$^{+26}$ ions.
Adopting $T_h=50$ keV, $T_c=0.001$ keV and $\eta=0.7$ we obtain 
$R_{c,h}\approx2\times10^3(M_c/M_h)$, which suggests that even a low-mass 
cool phase $M_c\sim 10^{-3}M_h$ could dominate the K$\alpha$ line emission. 
Keeping in mind that the
cool phase does not contribute to the X-ray continuum at 6.9 keV,
we thus conclude that the proposed mechanism could significantly increase
the equivalent width of the K$\alpha$ line compared to a single phase hot plasma. 
However, it is not clear whether the required penetration of Fe ions into the
cold gas could take place.
A magnetic field could inhibit the motion of Fe ions.
Also, if magnetic field effects are small, there is likely to be an intermediate temperature
zone between the hot and cool phases, and the Fe$^{+26}$ ions could recombine there.
Our main point is that the cool gas presents some possibilities for Fe line emission.
We argued in Section 3.1 that the absorption of the radio emission may also be related
cool gas that is mixed into the shocked, hot gas.

\section{DISCUSSION AND CONCLUSIONS}
\label{discussion}

The basic interpretation of the X-ray and radio data
 in terms of standard spherical models led
to inconsistencies.
One is that the
column density of matter to the X-ray emission is about
a factor 50 smaller than that needed to produce the X-ray luminosity.
Since the absorption column is $\propto \rho_w$ and the X-ray luminosity to
$\rho_w^2$, one possibility is that 
the luminosity is enhanced by clumpiness of the gas.
However, clumpiness would raise the shock 
velocity needed to produce a particular
X-ray temperature and the temperature is already close to that expected 
behind the
forward shock, so this solution does not seem feasible.
The situation, however, may be different in the case that the circumstellar clouds 
interact primarily with the cool dense shell. Then the cloud 
shock velocity is large enough to account for the high temperature 
derived from the observed X-ray spectrum.
The straightforward explanation is then that there is a global asymmetry in the
distribution of the circumstellar gas that allows a low column density in one direction,
while dense interaction is taking place over much of the rest of the solid angle (as
viewed from the supernova).
\cite{cd94} had suggested a scenario with equatorial mass loss to explain the 
presence of fast and intermediate velocity shock fronts in the Type IIn SN 1988Z.
Polarization observations of Type IIn supernovae have shown evidence for large
scale asymmetry, e.g., SN 1997eg \citep{hoffman08}, SN 1998S \citep{leonard00},
and SN 2010jl \citep{patat11}.

In this view, the column density to the radio emission would be small, as for the
X-ray emission, because they are both from the same region.
The external free-free absorption model for the radio absorption thus fails, and
an absorption mechanism internal to the emission is indicated.
We have argued that 
thermal absorption internal to the emitting region is a plausible
mechanism, as has previously been proposed for other
Type IIn supernovae \citep{weiler90}.

There is evidence that the X-ray properties we have found for SN 2006jd are not
unusual for a Type IIn event.
{\it Chandra} observations of SN 2001em and SN 2005dk have shown hard emission
\citep{pooley04,pooley07},
implying a high temperature.
In the case of SN 2001em, the column density is lower than expected from the
X-ray luminosity \citep{chugai06}, as in SN 2006jd.
Another point of comparison is the Type IIn SN 2010jl, which had an unabsorbed
$0.2-10$ keV luminosity of $\sim 7\times 10^{41}\ergs$ at ages of 2 months
and 1 yr \citep{chandra12}.
Although the X-ray luminosity is just a factor of 2 higher than that of SN 2006jd,
the absorbing column, $10^{24}$ cm$^{-2}$ at 2 months and
$3\times 10^{23}$ cm$^{-2}$ at 1 yr, is much higher.
In this case, the observed column is roughly consistent with that needed to
produce the X-ray luminosity.
The implication is that the observed line of sight passes through the dense
preshock gas for SN 2010jl, but not for SN 2006jd.

In this paper we have treated the radio and X-ray properties of SN 2006jd.
A more comprehensive study would include observations at other wavelengths.
SN 2006jd was detected as an infrared source at an age $\sim 1150$ days
\citep{fox11}.  Its luminosity was $\sim 10^{42}$ erg s$^{-1}$ which is
larger than the X-ray luminosity observed at this time.
If the interpretation of this emission as from circumstellar dust heated by shock
emission is correct \citep{fox11}, it appears that some of the X-ray emission
from the shocked region is shielded from the observer.
An infrared/X-ray ratio $>1$ has also been observed in other SNe IIn \citep{gfn+02}.

Our observations at both radio and X-ray wavelengths imply that the circumstellar density profile
($\rho\propto r^{-2}$) is flatter than the $s=2$ case that would be expected for a steady wind.
We find $s=(1.5-1.6)$; deviation from the steady case is plausible for a Type IIn supernova
because the mass loss may be due to an eruptive event.
In addition, the optical luminosity evolution of SN 2006jd shows a second peak at
$\sim 500$ days \citep{strit12} that is indicative of a nonstandard density distribution.

\acknowledgments
PC and RC are grateful to Craig Sarazin for discussions of X-ray emission and analysis.
The National Radio Astronomy Observatory is a facility of the National Science
Foundation operated under cooperative agreement by Associated Universities, Inc.
PC is supported by her NSERC Discovery grants.
RC and CI acknowledge support from NASA grants  NNX09AH58G (XMM) and GO9--0079X (Chandra), and  
NSF grant AST-0807727.

{\it Facilities:} \facility{CXO}, \facility{XMM-Newton}, \facility{Swift},
\facility{Very Large Array}, \facility{Expanded Very Large Array}, 
\facility{Giant Metrewave Radio Telescope}.

%



\tabletypesize{\scriptsize}
\begin{deluxetable}{cccccc}
\tablecaption{Radio observations of SN 2006jd
\label{tab:radio}}
\tablewidth{0pt}
\tablehead{
\colhead{Tel.} & \colhead{Date of} & \colhead{Days since} & \colhead{Freq.} & \colhead{Flux density} & \colhead{rms}\\
\colhead{} &\colhead{Observation} & \colhead{explosion} & \colhead{GHz} & \colhead{uJy} &
\colhead{uJy}
}
\startdata
VLA& 2007 Nov 21.28 &      404.74&   8.46 &   238  &   38     \\
VLA& 2007 Nov 26.38 &      409.84&   4.86 &   170  &   38     \\
VLA& 2007 Nov 26.38 &      409.84&   22.46&   324  &   54     \\
VLA& 2008 Mar 07.23 &      511.69&   8.46 &   305  &   46     \\
VLA& 2008 Mar 07.24 &      511.70&   4.86 &   232  &   37     \\
VLA& 2008 Mar 07.26 &      511.72&   22.46&   428  &   41     \\
VLA& 2008 Mar 07.30 &      511.76&   1.43 &   $<222$ &   74     \\
VLA& 2008 May 02.88 &      568.34&   43.31&   $<1680$&   560    \\
VLA& 2008 May 08.15 &      573.61&   8.46 &   525 &    50     \\
VLA& 2008 May 08.17 &      573.63&   4.86 &   480 &    54     \\
VLA& 2008 May 10.18 &      575.64&   1.43 &   $<291$&    97     \\
VLA& 2008 May 16.00 &      581.46&   8.46 &   509 &   42      \\
VLA& 2008 May 16.94 &      582.40&   1.43 &   209 &   63      \\
VLA& 2008 May 25.96 &      591.42&   22.46&   683 &   229     \\
VLA& 2008 Dec 15.38 &      794.84&   8.46 &   1024&   42      \\
VLA& 2008 Dec 15.40 &      794.86&   4.86 &   1035&   51      \\
VLA& 2008 Dec 15.41 &      794.87&   1.43 &   250 &   65      \\
VLA& 2008 Dec 15.45 &      794.91&  22.46 &   624 &   90      \\
VLA& 2009 Feb 03.16 &      844.62&   8.46 &   1033&   23      \\
VLA& 2009 Feb 03.22 &      844.68&   4.86 &   1192&   26      \\
VLA& 2009 Feb 03.28 &      844.74&   1.43 &   247 &   43      \\
VLA& 2009 Feb 03.34 &      844.80&   22.46&   665 &   43      \\
VLA& 2009 Apr 02.02 &      902.48&   8.46 &   1069&   71      \\
VLA& 2009 Apr 02.04 &      902.50&   4.86 &   1178&   55      \\
VLA& 2009 Apr 02.06 &      902.52&   1.43 &   552 &   47      \\
VLA& 2009 Apr 02.10 &      902.56&   22.46&   390 &   55      \\
VLA& 2009 Aug 07.67 &     1030.13&   8.46 &   770 &   50      \\
VLA& 2009 Aug 07.69 &     1030.15&   4.86 &   1608&   52      \\
VLA& 2009 Aug 07.71 &     1030.17&   1.43 &   743 &   123     \\
VLA& 2009 Aug 07.75 &     1030.21&  22.46 &   $<645$&   215     \\
GMRT& 2009 Oct 21.00 &     1104.46&  1.28  &   730 &  124 \\
GMRT& 2010 Feb 09.65 &     1216.11&  1.28  &   674 & 79 \\
GMRT& 2010 Apr 24.52 &     1289.98&  0.61  &   $<441$ & 147   \\
EVLA& 2010 Jun 06.91 &     1333.28    &  4.96  &   2216 & 54 \\
EVLA& 2010 Jul 25.67 &     1382.13   &  8.46 &  918 & 37\\ 
EVLA& 2011 Jul 09.72 & 1731.18 & 4.50 & 1659 & 43 \\
EVLA & 2011 Jul 09.72 & 1731.18 & 7.92 & 1045 & 33\\
EVLA & 2011 Jul 30.65 & 1752.11 &22.46 &$<405$& 135\\
EVLA & 2011 Jul 31.72 & 1753.18 & 1.39 & 1642 & 96\\
EVLA& 2012 Apr 01.96 & 1998.42 & 4.50 & 1385 & 350 \\
EVLA& 2012 Apr 01.96 & 1998.42 & 7.90 & 966 & 25 \\
EVLA & 2012 Apr 03.02 & 1999.48 & 1.39 & 1564 & 208\\
EVLA & 2012 Apr 07.99 & 2004.45 &22.46 &$468$& 66
\enddata
\tablecomments{Here the upper limits are 3$-\sigma$ upper limits.}
\end{deluxetable}

\tabletypesize{\scriptsize}
\begin{deluxetable}{lcccc}
\tablecaption{Details of X-ray observations of SN 2006jd
\label{tab:xray}}
\tablewidth{0pt}
\tablehead{
\colhead{Date of} & \colhead{Mission} & \colhead{Inst.} & \colhead{Obs.} & \colhead{Exposure} \\
\colhead{Observation} &\colhead{} & \colhead{} & \colhead{ID} & \colhead{ks}
}
\startdata
2007 Nov 16.73 & \emph{Swift} & XRT & 00031016001 & 2.11\\
2007 Nov 21.49 & \emph{Swift} & XRT & 00031016002 & 2.92\\
2007 Nov 22.03 & \emph{Swift} & XRT & 00031016003 & 1.61\\
2007 Nov 22.84 & \emph{Swift} & XRT & 00031016004 & 3.06\\
2007 Dec 18.01    & \emph{Swift} & XRT & 00031016006 & 9.55 \\
2008 Jan 15.32 &  \emph{Swift} & XRT & 00031016007 & 9.26 \\
2008 Feb 20.18 & \emph{Swift} & XRT & 00031016008 & 3.93 \\
2008 Feb 21.32 & \emph{Swift} & XRT & 00031016009 & 6.06 \\
2008 Sep 09.02 & \emph{Swift} & XRT & 00031016010 & 1.76\\
2008 Sep 10.82 & \emph{Swift} & XRT &00031016011 & 2.49\\
2008 Sep 11.02 &  \emph{Swift} & XRT & 00031016012 & 7.36 \\
2009 Apr 07.20    &    \emph{XMM} & EPIC-PN & 0550850101 & 42.92\\
2009 Sep 10.12    & \emph{Swift} & XRT & 00031016013 & 9.79 \\
2009 Sep 14.00  	& \emph{Chandra} & ACIS-S & 10127 & 37.24\\
2011 Mar 08.32 & \emph{Swift} & XRT & 00037767006 & 1.06\\
2011 Mar 08.38 & \emph{Swift} & XRT & 00037767004 & 0.99\\
2011 Mar 08.46 & \emph{Swift} & XRT & 00037767005 & 1.01\\
2011 Mar 09.32 &\emph{Swift} & XRT & 00037767007 & 3.72\\
2011 Mar 10.81 & \emph{Swift} & XRT & 00037767008 & 3.12\\
2011 Mar 12.41 & \emph{Swift} & XRT & 00037767009 & 2.35\\
\enddata
\end{deluxetable}

\begin{deluxetable}{lccccc}
\tablecaption{Count rates obtained from \emph{Swift}-XRT Observations
\label{tab:swift}}
\tablewidth{0pt}
\tablehead{
\colhead{ObsID} &   \colhead{Exposure} &   \colhead{RA} &   \colhead{Decl.}
&  \colhead{Error}  & \colhead{Count Rate}   \\
\colhead{} &   \colhead{s} &   \colhead{J2000} &   \colhead{J2000}      &  \colhead{$\arcsec$}
  & \colhead{cts/s}   
}
\startdata
0031016001-004&    9695.659 &        08:02:07.2 &     +00:48:31.9&     3.47  
& $ (6.816\pm1.100)\times10^{-03}$\\   
0031016006    &  9547.694   &     08:02:07.1    &  +00:48:34.0   &  5.55    
&  $(6.464\pm1.100)\times10^{-03}$\\    
0031016007    &  9261.903   &     08:02:07.3    &  +00:48:33.7   &  2.94   
&   $(7.421\pm1.200)\times10^{-03}$\\    
0031016008-009 &    9991.502 &       08:02:07.5 &     +00:48.33.3&     2.00
&    $(6.011\pm1.000)\times10^{-03}$ \\   
00310160010-012 &  11621.214  &     08:02:07.1  &    +00:48:28.9 &    5.59
&    $(8.094\pm1.100)\times10^{-03}$   \\ 
0031016013  &    9750.140     &   08:02:07.4    &  +00:48:29.5   &  2.05 
&     $(4.549\pm0.910)\times10^{-03}$ \\ 
0037767004-009 &    12151.953  &     08:02:07.5 &     +00:48:30.4&     1.52
&   $(5.608\pm0.910)\times10^{-03}$ \\    
\enddata
\end{deluxetable}

\begin{deluxetable}{lcccccc}
\tabletypesize{\tiny}
\tablecaption{Model fits to SN 2006jd radio data (assuming $\alpha=-1.04$)
\label{tab:radiofit}}
\tablewidth{0pt}
\tablehead{
\colhead{Model} & \colhead{${\chi}^2/{\nu}$}  &
 \colhead{Param-1 } & \colhead{Param-2 } &   \colhead{Param-3 } &  \colhead{Param-4} &  \colhead{Param-5 } \\
} 
\startdata 
External FFA ($s=2$) & 2.41 & $K_1=(2.27\pm0.03)\times10^3$ & $K_2=0.46\pm0.01$ & $\delta=3.46\pm0.02$ & $\cdots$ & $\cdots$\\
External FFA (flatter $s$) & 2.11 & $K_1=(2.05\pm0.03)\times10^3$ & $K_2=0.40\pm0.01$ & $s=1.77\pm0.02$ &  $m=1.09\pm0.01$ & $\cdots$\\
Internal FFA & 1.26 & $K_1=(2.30\pm0.04)\times10^3$ & $K_3=1.20\pm0.04$ & $\beta=0.46\pm0.04$ & $\delta'=3.15\pm0.07$ & $\cdots$\\
Internal+External FFA & 1.30 & $K_1=(2.31\pm0.04)\times10^3$ &  $K_2=0.002\pm0.003$ & $K_3=1.22\pm0.05$ & $\delta=3.58\pm0.04$ & $\delta'=3.16\pm0.07$
\enddata
\tablecomments{see}
\end{deluxetable}


\begin{deluxetable}{lccccccc}
\tabletypesize{\tiny}
\tablecaption{Spectral model fits to the \emph{XMM-Newton} spectrum
\label{tab:xspec-xmm}}
\tablewidth{0pt}
\tablehead{
\colhead{Model} & \colhead{${\chi}^2/{\nu}$}  &
\colhead{N$_{\rm H}$}  & \colhead{Param-1 } & \colhead{Param-2 } & \colhead{Abs. flux} & \colhead{Unabs. flux} & \colhead{Total flux} \\
} 
\startdata 
{\bf PowerLaw} & 1.03(123) & $1.41^{+0.37}_{-0.30} \times10^{21}$
& $\Gamma=1.31^{+0.15}_{-0.11}$ & $\cdots$ & $2.90\times10^{-13}$ &
$3.33\times10^{-13}$ &  $3.78\times10^{-13}$\\
 ~~+ Gaussian & $\cdots$ & $\cdots$  
& $E=6.76^{+0.10}_{-0.10}$ & $EW=1.02^{+0.16}_{-0.16}$ & $4.57\times10^{-14}$ &  $4.58\times10^{-14}$ & 
$\cdots$\\ 
{\bf NEI} & 1.09(125) & $1.22^{+0.24}_{-0.18} \times10^{21}$
& $kT=79.65^{+\cdots}_{-48.02}$ & $\tau=6.28^{+2.66}_{-1.55} \times10^{11}$ & 
$3.25\times10^{-13}$ & $3.59\times10^{-13}$ &
$3.59\times10^{-13}$\\
{\bf Mekal} & 1.12(126) & $1.38^{+0.22}_{-0.21} \times10^{21}$ 
& $kT=36.77^{+\cdots}_{-20.18}$ & $\cdots$ &
$3.09\times10^{-13}$ & $3.48\times10^{-13}$ 
& $3.48\times10^{-13}$\\
{\bf Bremsstrahlung} & 1.02(123) & $1.25^{+0.21}_{-0.23} \times10^{21}$ &
$kT=56.76^{+56.80}_{-28.83}$ & $\cdots$ &
$2.91\times10^{-13}$ & $3.26\times10^{-13}$
& $3.64\times10^{-13}$\\
 ~~+ Gaussian & $\cdots$ & $\cdots$
& $E=6.76^{+0.11}_{-0.10}$ & $EW=1.38^{+0.18}_{-0.13}$ & $3.75\times10^{-14}$ &  $3.76\times10^{-14}$ &
$\cdots$
\enddata
\tablecomments{Here $N_H$ is in cm$^{-2}$, $E$ and $EW$ are in keV,
$\tau$ in cm$^{-3}$ s,
and the fluxes are in
erg cm$^{-2}$ s$^{-1}$. The fluxes  mentioned are in  the
$0.2-10.0$~keV energy range. The abs.\ and unabs.\ fluxes are
of that particular component in the model,
whereas the total flux is for the total  unabsorbed flux of the model.}
\end{deluxetable}

\begin{deluxetable}{lccccccc}
\tabletypesize{\tiny}
\tablecaption{Spectral model fits to the \emph{Chandra} spectrum
\label{tab:xspec-chandra}}
\tablewidth{0pt}
\tablehead{
\colhead{Model} & \colhead{${\chi}^2/{\nu}$}  &
\colhead{N$_{\rm H}$}  & \colhead{Param-1 } & \colhead{Param-2 } & \colhead{Abs. flux} & \colhead{Unabs. flux} & \colhead{Total flux} \\
} 
\startdata 
{\bf PowerLaw} & 0.90(48) & $1.83^{+0.61}_{-0.53} \times10^{21}$
& $\Gamma=1.34^{+0.17}_{-0.16}$ & $\cdots$ & $2.82\times10^{-13}$ &
$3.24\times10^{-13}$ &  $3.44\times10^{-13}$\\
 ~~+ Gaussian & $\cdots$ & $\cdots$  
& $E=6.67^{+1.24}_{-0.21}$ & $EW=0.73^{+1.12}_{-\cdots}$ & $1.98\times10^{-14}$ &  $1.98\times10^{-14}$ & 
$\cdots$\\ 
{\bf NEI} & 0.83(50) & $1.59^{+0.46}_{-0.38} \times10^{21}$
& $kT=35.93^{+\cdots}_{-20.32}$ & $\tau=3.72^{+4.51}_{-0.80} \times10^{11}$ & 
$3.02\times10^{-13}$ & $3.36\times10^{-13}$ &
$3.36\times10^{-13}$\\
{\bf Mekal} & 0.82(51) & $1.80^{+0.39}_{-0.35} \times10^{21}$ 
& $kT=20.67^{+35.71}_{-7.63}$ & $\cdots$ &
$2.96\times10^{-13}$ & $3.36\times10^{-13}$ 
& $3.36\times10^{-13}$\\
{\bf Bremsstrahlung} & 0.91(48) & $1.61^{+0.47}_{-0.35} \times10^{21}$ &
$kT=41.61^{+\cdots}_{-27.77}$ & $\cdots$ &
$2.78\times10^{-13}$ & $3.13\times10^{-13}$
& $3.35\times10^{-13}$\\
 ~~+ Gaussian & $\cdots$ & $\cdots$
& $E=6.72^{+0.74}_{-0.22}$ & $EW=0.83^{+0.35}_{-\cdots}$ & $2.20\times10^{-14}$ &  $2.20\times10^{-14}$ &
$\cdots$
\enddata
\tablecomments{Here $N_H$ is in cm$^{-2}$, $E$ and $EW$ are in keV,
$\tau$ in cm$^{-3}$ s,
and the fluxes are in
erg cm$^{-2}$ s$^{-1}$. The fluxes  mentioned are in the
0.2--10.0~keV energy range. The abs.\ and unabs.\ fluxes are
for that particular component in the model,
whereas the total flux is for the total  unabsorbed flux of the model.}
\end{deluxetable}

\begin{deluxetable}{ccccc}
\tablecaption{Unabsorbed $0.2-10$ keV X-ray fluxes of SN 2006jd at various epochs
\label{tab:flux}}
\tablewidth{0pt}
\tablehead{
\colhead{Date of} & \colhead{Days since} & \colhead{Instrument} & 
\colhead{Flux} &
\colhead{Luminosity}\\
\colhead{Observation} & \colhead{Explosion} & \colhead{}& \colhead{$10^{-13}$erg/cm$^2$/s} 
&\colhead{$10^{41}$erg/s}
}
\startdata
2007 Nov 16.73--22.84   & $403.2\pm3.1$ & \emph{Swift}-XRT &  $4.51\pm0.73$
& $3.37\pm0.54$\\
2007 Dec 18.01          & 431.5	& \emph{Swift}-XRT &  $4.28\pm0.73$
& $3.19\pm0.54$\\
2008 Jan 15.32		& 459.8	& \emph{Swift}-XRT & $4.91\pm0.79$
& $3.67\pm0.59$\\	
2008 Feb 20.18--21.32 	& $496.2\pm0.6$&\emph{Swift}-XRT & $3.98\pm0.66$
& $2.97\pm0.49$\\    
2008 Sep 09.02--11.02	& $698.5\pm1.0$&\emph{Swift}-XRT & $5.35\pm0.73$
& $4.00\pm0.54$\\    
2009 Apr 07.20		& 907.7	&\emph{XMM} EPIC-PN& $3.63^{+0.33}_{-0.30}$
& $2.70^{+0.25}_{-0.23}$\\    
2009 Sep 10.12		& 1063.6	& \emph{Swift}-XRT & $3.01\pm0.60$
& $2.25\pm0.45$\\    
2009 Sep 14.00		& 1067.5	& \emph{Chandra} ACIS-S& $3.38^{+0.89}_{-0.93}$
& $2.52^{+0.70}_{-0.66}$\\    
2011 Mar 08.32--12.41	& $1609.8\pm2.0$ & \emph{Swift}-XRT & $3.71\pm0.60$
& $2.77\pm0.45$
\enddata
\end{deluxetable}



\end{document}